\documentclass[prd,onecolumn,showpacs,preprintnumbers,amsmath,amssymb,floatfix,nofootinbib]{revtex4}

\usepackage{graphicx}
\usepackage{dcolumn}
\usepackage{xspace}
\usepackage[utf8]{inputenc}
\usepackage{float}
\usepackage{subfigure}
\usepackage{url}

\usepackage{color}
\usepackage{comment}
\usepackage{bbm}

\definecolor{BlueViolet}{rgb}{0.2, 0.00, 0.7}
\definecolor{Blue}{rgb}{0.15, 0.00, 0.9}
\definecolor{halayaube}{rgb}{0.4, 0.22, 0.33}
\definecolor{sanddune}{rgb}{0.59, 0.44, 0.09}
\usepackage[colorlinks=true,linkcolor=Blue,citecolor=Blue,urlcolor=BlueViolet,
hyperfootnotes=false]{hyperref}

\begin{document}

\preprint{}
\preprint{}
\title{Nucleon resonances with spin $\frac{3}{2}$ and isospin $\frac{1}{2}$}

\author{K.~Azizi}
\thanks{Corresponding author}
\affiliation{Department of Physics, Do\u gu\c s University,
Ac{\i}badem-Kad{\i}k\"oy, 34722 Istanbul, Turkey}
\affiliation{Department of Physics, University of Tehran, North Karegar Avenue, Tehran
14395-547, Iran}
\author{Y.~Sarac}
\affiliation{Electrical and Electronics Engineering Department,
Atilim University, 06836 Ankara, Turkey}
\author{H.~Sundu}
\affiliation{Department of Physics, Kocaeli University, 41380 Izmit, Turkey}

\date{\today}

\begin{abstract}
Investigation of the nucleon's excited states has always become an important research topic because of the rich information they provide. Since their first observation, dating back about 70 years, the investigation of their various parameters contributed both to the development of the quark model and a better understanding of the QCD as the theory of strong interaction.  Their investigation still has importance. The researches conducted on the nucleon excited states are helpful to probe the missing resonances predicted by the quark model but not observed yet. With this motivation, we study the low lying nucleon resonance with $I(J^P)=\frac{1}{2}(\frac{3}{2}^{-})$ and its corresponding orbital and radial excitations with $I(J^P)=\frac{1}{2}(\frac{3}{2}^+)$ and $I(J^P)=\frac{1}{2}(\frac{3}{2}^-)$, respectively. Using the QCD sum rule method, we calculate the masses and pole residues of these states. The obtained mass results are
consistent with the mass ranges presented in PDG for the resonances $N(1520)(3/2^-)$, $N(1700)(3/2^-)$, and $N(1720)(3/2^+)$. The results of masses and residues of these states may be used as input parameters to calculate various quantities related to their electromagnetic, weak and strong interactions with other particles with the aim of getting more information on their natures and structures.
\end{abstract}
\maketitle

\vspace{-1mm}
\maketitle
\renewcommand{\thefootnote}{\#\arabic{footnote}}
\setcounter{footnote}{0}
\section{\label{sec:level1}Introduction}

The ongoing progress in experimental studies brings us new data related to not only conventional states but also nonconventional ones. Among the collected data are the ones belonging to the light and heavy baryons and their excited states with increasing confidence levels. The observations of their excited states have increased the numbers of these baryons enriching our understanding of the strong interaction. The discrepancy between the number of experimentally observed nucleon excited states and the expectation of the quark model makes the subject more intriguing. Therefore understanding their spectroscopic parameters, substructures and interactions has importance to better understand the strong interaction, and the missing resonances as well. Investigation of the spectroscopy of these states may help us improve our comprehension of the confinement and interaction mechanism of the quarks and gluons in the nonperturbative domain of QCD and thus provides better understanding of the strong interaction.

Until now, there have been a large number of experimental and theoretical studies on the excited states of the nucleon. These excited states were firstly observed in $\pi N$ scattering. Their existence and properties such as masses and decay widths were obtained mostly from partial wave analysis of $\pi N$ elastic scattering. In this respect, the Karlsruhe-Helsinki (KH80)~\cite{KH80}, Carnegie-Mellon$-$Berkeley (CMB80)~\cite{CMB80}, and George Washington U (GWU)~\cite{CWU} groups provided the most comprehensive analyses. Though many states were identified by $\pi N$ elastic scattering, their numbers were less than that were predicted by the standard quark model. Since this non-observation could be a result of the weak coupling of these states to $\pi N$, besides the $\pi N$ scattering, these states and their properties have been searched through experimental investigations based on various other interaction mechanisms such as photoproduction and electroproduction. Some of these experimental researches focusing on nucleon resonances were conducted at Jefferson Lab, the Electron Stretcher Accelerator (ELSA), the Mainz Microtron (MAMI), and the
Grenoble Anneau Accelerateur Laser (GRAAL) facility, Super Photon Ring (SPring-8)~\cite{Aznauryan:2012ba,Kashevarov:2012wy,Ass03,Crede:2013sze,Gutz:2014wit,So17,Ann15,Bu16,Leps,Kl17,Be17}. 

The recent observations of possible excited states of light and heavy baryons~\cite{Yelton:2018mag,Aaij:2017nav,Aaij:2018tnn,Aaij:2018yqz} have recollected the attentions on these excited states, and several analyses on these states containing both negative and positive parity excitations were performed (see for instance the Refs~\cite{Aliev:2018yjo,Aliev:2018vye,Aliev:2018lcs} and the references therein). With these improvements not only the excited states of the baryons containing heavy quarks but also those containing only light quarks have attracted interests. In Particle Data Group (PDG) listing there exist thirteen $N^*$ resonances with status $4$ stars~\cite{Tanabashi18}. On the other hand, the number of states predicted by the quark model at the same energy region is larger~\cite{Capstick:2000qj}. The reason behind these may be attributed to the effective degrees of freedom used in the model. Therefore to interpret their internal structures different theoretical models were considered~\cite{Aznauryan:2012ba,Capstick:2000qj,Aznauryan:2011qj}. The missing $N^*$ states problem may also be a result of the weak coupling of these states to the $\pi N$ states or other observation channels. The properties of the excited baryons were investigated via various theoretical models such as quark models~\cite{Capstick:2000qj,Valcarce:2005rr,NIsgur,Aznauryan:2007ja,Aznauryan:2012ec,NIsgur86,LYaGlozman,MGiannini,GGalata,DeSanctis:2014ria,Eichmann:2016nsu,Ramalho:2016buz,Shah:2018ont}, lattice QCD~\cite{WalkerLoud:2008bp,Lin:2008pr,Edwards:2011jj}, covariant three-body Faddeev approach~\cite{Sanchis-Alepuz:2014sca} and basis light front quantization approach~\cite{JPVary}. Masses and other properties of negative parity nucleon states were studied using QCD sum rule method~\cite{Jido:1996ia,Oka:1996zz,Lee:1998cx,YKondo,Braun:2014wpa,Aliev:2014foa,Azizi:2015fqa,Aliev:2019tmk,Azizi:2015jya}, as well. 

From the PDG listing of nucleon states it can be seen that there are many nucleon resonances as previously mentioned. Among these states are $N(1440)(1/2^+)$, $N(1520)(3/2^-)$, $N(1535)(1/2^-)$, $N(1650)(1/2^-)$, $N(1675)(5/2^-)$, $ N(1680)(5/2^+)$,  $N(1710)(1/2^+)$, $N(1720)(3/2^+)$, $N(1895)(1/2^-)$, $N(1900)(3/2^+)$, $N(2190)(7/2^-)$, $N(2220)(9/2^+)$ and $N(2250)(9/2^-)$ with 4-star status~\cite{Tanabashi18}. In this work, our aim is to consider the nucleon resonances $N(1520)(3/2^-)$ and $N(1720)(3/2^+)$ with $I(J^P)=\frac{1}{2}(\frac{3}{2}^{-})$ and $I(J^P)=\frac{1}{2}(\frac{3}{2}^{+})$ quantum numbers, respectively. Beside these two 4-star status resonances we also take into account the $N(1700)(3/2^-)$state  with  $I(J^P)=\frac{1}{2}(\frac{3}{2}^{-})$, which has a status 3 stars in PDG. To obtain a deeper understanding on their structures we investigate their mass spectrum. Using proper interpolating current carrying the same quantum numbers with the considered states we calculate their masses considering them as ground, and low lying orbital and radial excited states. In the calculation, we use the QCD sum rules method~\cite{Shifman:1978bx,Shifman:1978by,Ioffe81}, which is among the successful nonperturbative methods applied in many studies resulting in reliable predictions, so far.  Such analyses improve our understanding of the interaction mechanism of the quarks in the nonperturbative regime of QCD and provide us a better understanding of the strong interaction in the low energy domain of the QCD and the confinement. This may also improve our understanding about all the other $N^*$ states which are both observed and not yet observed. Better understanding of the dynamics related with three-quark systems and their arrangement may shed light on the unobserved states. By means of such investigation, we may also test the findings of different approaches and the experiments.

The outline of the article is as follows: In Section II we present the details of our QCD sum rules calculations. In Section III the numerical analyses of the results are given. Section IV is devoted to the summary and conclusion.


\section{Sum Rules Calculations}

To extract the masses of the baryons considered in the present work the following two point correlation function is used:
\begin{equation}
\Pi _{\mu \nu}(q)=i\int d^{4}xe^{iq\cdot x}\langle 0|\mathcal{T}\{J_{\mu}(x)J^{\dagger}_{\nu}(0)\}|0\rangle ,  \label{eq:CorrF1}
\end{equation}        
where $J_{\mu}$ is the interpolating current that carries the quantum numbers of the considered baryons. In this work the mentioned baryons are $N(1520)3/2^-$,  $N(1700)3/2^-$ and $N(1720)3/2^+$. For these baryons the interpolating current that can create or annihilate them has the following form~\cite{Lee:2002jb}
\begin{eqnarray}
J_{\mu}=\epsilon_{abc}[(u^{aT}C\sigma_{\alpha\beta}d^b)\sigma^{\alpha\beta}\gamma_{\mu}u^c-(u^{aT}C\sigma_{\alpha\beta}u^b)\sigma^{\alpha\beta}\gamma_{\mu}d^c],
\end{eqnarray} 
in which $a$, $b$ and $c$ represent the color indices, $C$ is the charge conjugation operator and $T$ represents transpose. Note that the above current annihilates or creates both the ground state nucleon resonance and its excited states having both positive and negative parities. Moreover, it couples not only to spin-$\frac{3}{2}$ states but also to spin-$\frac{1}{2}$ states. Therefore in the analyses one needs to remove these unwanted contributions coming from spin-$\frac{1}{2}$ states. These can be done by proper choices of the Lorentz structures used in the analyses that do not include the spin-$\frac{1}{2}$ pollution.

The correlator, Eq.~(\ref{eq:CorrF1}), can be calculated either in terms of hadronic degrees of freedom such as mass of the hadron and its pole residue or in terms of the QCD degrees of freedom such as masses of the quarks, quark-gluon condensates etc. The QCD sum rules are attained after the calculations of both and matching them via dispersion relation considering coefficients of the same Lorentz structures obtained from each side. We also apply Borel transformation to both sides and this provides exponential suppression on the higher states and continuum and factorial suppression on the terms having higher dimensional operators.   

To calculate the correlator in terms of hadronic degrees of freedom, it is saturated by complete sets of hadronic states having the same quantum numbers with the chosen interpolating current. This results in
\begin{eqnarray}
\Pi_{\mu\nu}^{\mathrm{Had}}(q)&=&\frac{\langle 0|J_{\mu} |N^*(q,s)\rangle \langle N^*(q,s)|J^\dagger_{\nu}|0\rangle}{q^{2}-m^{2}}
+\frac{\langle 0|J_{\mu} |N_1^*(q,s)\rangle \langle N_1^*(q,s)|J^\dagger_{\nu}|0\rangle}{q^{2}-m_{1}^{2}}+\frac{\langle 0|J_{\mu} |N_2^*(q,s)\rangle \langle N_2^*(q,s)|J^\dagger_{\nu}|0\rangle}{q^{2}-m_{2}^{2}}
\nonumber\\
&+&\ldots,
\label{eq:phys}
\end{eqnarray}       
The terms given in Eq.~(\ref{eq:phys}) correspond to the contributions of the ground state $N^*$ spin-$\frac{3}{2}$ baryon with negative parity and its excitations with negative and positive parities, respectively. Their one-particle states are represented by $ |N^*(q,s)\rangle$, $ |N_1^*(q,s)\rangle$ and $ |N_2^*(q,s)\rangle$ respectively and $m$, $m_1$ and $m_2$ are their corresponding masses. The contributions of the higher states and continuum are denoted by the $\ldots$ . The matrix elements in above equation are parameterized in terms of the pole residues as follows
\begin{eqnarray}
\langle 0|J_{\mu } |N^{*}(q,s)\rangle &=&\lambda^{*}u_{\mu}(q,s),
\nonumber \\
\langle 0|J_{\mu } |N_1^{*}(q,s)\rangle
 &=&\lambda_1^{*}u_{\mu}(q,s),
\nonumber \\
\langle 0|J_{\mu } |N_{2}^{*}(q,s)\rangle
 &=&\lambda_2^{*}\gamma_5u_{\mu}(q,s),
\label{eq:Res2}
\end{eqnarray}
where $u_{\mu}$ is the spin-vector in Rarita Schwinger representation. In the calculation we need the summation over spin given by
\begin{eqnarray}\label{Rarita}
\sum_s  u_{\mu} (q,s)  \bar{u}_{\nu} (q,s) &= &-(\!\not\!{q} + m)\Big[g_{\mu\nu} -\frac{1}{3} \gamma_{\mu} \gamma_{\nu} - \frac{2q_{\mu}q_{\nu}}{3m^{2}} +\frac{q_{\mu}\gamma_{\nu}-q_{\nu}\gamma_{\mu}}{3m} \Big].
\end{eqnarray}
As we mentioned our current also couples to spin-$\frac{1}{2}$ states with positive and negative parities and corresponding matrix elements are given as 
\begin{eqnarray}
\langle 0|J_{\mu}|\frac{1}{2}^+(q)\rangle =A_{\frac{1}{2}^+}(\gamma_{\mu}+\frac{4q_{\mu}}{m_{\frac{1}{2}^+}})\gamma_5 u(q,s),
\end{eqnarray}
and
\begin{eqnarray}
\langle 0|J_{\mu}|\frac{1}{2}^-(q)\rangle =A_{\frac{1}{2}^-}(\gamma_{\mu}+\frac{4q_{\mu}}{m_{\frac{1}{2}^-}})u(q,s),
\end{eqnarray}
respectively. This matrix elements indicate that the terms in the calculation that are proportional to $\gamma_{\mu}$ and $q_{\mu}$ include contributions from spin-$\frac{1}{2}$ states. Their unwanted pollution can be avoided with a proper choice of the Lorentz structures containing contributions only from spin-$\frac{3}{2}$ states. With this aim, we select the structures $g_{\mu\nu}$ and $\not\!q g_{\mu\nu}$. From the calculation of the hadronic side the following result is obtained:
\begin{eqnarray}\label{PhyssSide}
\Pi_{\mu\nu}^{\mathrm{Had}}(q)&=&-\frac{\lambda^{*}{}^{2}}{q^{2}-m^{2}}(\!\not\!{q} + m)\Big[g_{\mu\nu} -\frac{1}{3} \gamma_{\mu} \gamma_{\nu} - \frac{2q_{\mu}q_{\nu}}{3m^{2}} +\frac{q_{\mu}\gamma_{\nu}-q_{\nu}\gamma_{\mu}}{3m} \Big]\nonumber \\
&-&\frac{\lambda_1^{*}{}^{2}}{q^{2}-m_1{}^{2}}(\!\not\!{q} + m_1)\Big[g_{\mu\nu} -\frac{1}{3} \gamma_{\mu} \gamma_{\nu} - \frac{2q_{\mu}q_{\nu}}{3m_1^{2}} +\frac{q_{\mu}\gamma_{\nu}-q_{\nu}\gamma_{\mu}}{3m_{1}} \Big]\nonumber\\
&-&\frac{\lambda_2^{*}{}^{2}}{q^{2}-m_2^{2}}(\!\not\!{q} - m_2)\Big[g_{\mu\nu} -\frac{1}{3} \gamma_{\mu} \gamma_{\nu} - \frac{2q_{\mu}q_{\nu}}{3m_2^{2}} +\frac{q_{\mu}\gamma_{\nu}-q_{\nu}\gamma_{\mu}}{3m_2} \Big]+\ldots,
\end{eqnarray}
and considering the mentioned Lorentz structures, $g_{\mu\nu}$ and $\not\!q g_{\mu\nu}$, the result becomes
\begin{eqnarray}
\Pi _{\mu \nu}^{\mathrm{Had}}(q)&=&-\frac{\lambda^{*}{}^2}{q^{2}-m^{2}} \left( \!\not\!{q}g_{\mu\nu}+m g_{\mu\nu} \right)
-
\frac{\lambda_{1}^{*}{}^2}{q^{2}-m_{1}^{2}} \left( \!\not\!{q}g_{\mu\nu}+m_{1}g_{\mu\nu}\right) \nonumber\\
&-&\frac{\lambda_2^{*}{}^2}{q^{2}-m_2^{2}} \left( \!\not\!{q}g_{\mu\nu}-m_2 g_{\mu\nu} \right)+\ldots.
\label{eq:CorFun1}
\end{eqnarray}
Final form of the Eq.~(\ref{eq:CorFun1}) is obtained after Borel transformation. This transformation is applied to suppress the contributions of the higher states and continuum. After Borel transformation with respect to $-q^2$ the following result is obtained for the hadronic side
\begin{eqnarray}
\mathcal{\widehat B}\Pi _{\mu \nu}^{\mathrm{Had}}(q)&=&\lambda^{*}{}^2 e^{-\frac{m^{2}}{M^{2}}} \left( \!\not\!{q}g_{\mu\nu}+mg_{\mu\nu} \right)
+
\lambda_{1}^*{}^2 e^{-\frac{m_{1}^{2}}{M^{2}}} \left( \!\not\!{q}g_{\mu\nu}+m_{1}g_{\mu\nu}\right) \nonumber\\
&+&\lambda_2^{*}{}^2 e^{-\frac{m_2^{2}}{M^{2}}} \left( \!\not\!{q}g_{\mu\nu}-m_{2}g_{\mu\nu} \right)+\cdots,\nonumber\\
\label{eq:CorFunBorel}
\end{eqnarray}
and, since we will obtain independent sum rules corresponding to each structure present in this result, we express the coefficient of $\not\!q g_{\mu\nu}$ structure obtained from above equation as $\Pi_1$ and that of $g_{\mu\nu}$ as $\Pi_2$ hereafter. These are the structures that we use in the analyses.

The QCD side of the calculations is obtained using the interpolating current inside the correlator explicitly  and calculate it via the operator product expansion. After insertion of the interpolating current into Eq.~(\ref{eq:CorrF1}) and making the possible contractions between the quark fields via Wick's theorem, the correlator leads to an  expression containing the light quark propagators. In the calculations we use the light quark propagator explicitly in $x$ space which has the following form
\begin{eqnarray}
 S_{q}^{ab}(x)&=&i\frac{x\!\!\!/}{2\pi^{2}x^{4}}\delta_{ab}-\frac{m_{q}}{4\pi^{2}x^{2}}\delta_{ab}-\frac{\langle
 \overline{q}q\rangle}{12}\Big(1-i\frac{m_{q}}{4}x\!\!\!/\Big)\delta_{ab}-\frac{x^{2}}{192}m_{0}^{2}\langle
 \overline{q}q\rangle\Big( 1-i\frac{m_{q}}{6}x\!\!\!/\Big)\delta_{ab}-\frac{ig_{s}G_{ab}^{\theta\eta}}{32\pi^{2}x^{2}}\Big[x\!\!\!/\sigma_{\theta\eta} +\sigma_{\theta\eta}x\!\!\!/ \Big]
 \nonumber\\&-&\frac{x\!\!\!/ x^{2}g_s^2}{7776}\langle
 \overline{q}q\rangle^2\delta_{ab}-\frac{x^4\langle
 \overline{q}q\rangle\langle
 g_s^2G^2\rangle}{27648}\delta_{ab}+\frac{m_q}{32\pi^2}[ln(\frac{-x^2\Lambda^2}{4})+2 \gamma_E]g_{s}G_{ab}^{\theta\eta}\sigma_{\theta\eta}+\cdots,
\end{eqnarray}
where the $\gamma_E$ is the Euler constant, $\gamma_E \simeq 0.577$, and $\Lambda$ is the QCD scale parameter. After inserting this propagator in the places of the propagators, we make the Fourier and Borel transformations and apply continuum subtraction. All this calculations result in the following expressions:
\begin{eqnarray}
\mathcal{\widehat B}\Pi_{\mu \nu,i}^{\mathrm{QCD}}&=&\int_0^{s_0} e^{-\frac{s}{M^2}}\rho_{\mu\nu,i}(s) ds+\Gamma_{\mu\nu,i},
\end{eqnarray}
where subindex $i=1,2$ is used to refer the structures $\not\!q g_{\mu\nu}$ and $g_{\mu\nu}$, respectively, and the $\rho(s)_{\mu\nu,i}$ and $\Gamma_{\mu\nu,i}$ are obtained as follows
\begin{eqnarray}
\rho_{\mu\nu,1}(s)&=&\frac{1}{120 \pi^4}  (25 \langle g^2G^2\rangle - 18 s^2),\nonumber\\
\rho_{\mu\nu,2}(s)&=&\frac{2}{3 \pi^2}  \Big[2 s (\langle \bar{d}d\rangle - 4 \langle \bar{u}u\rangle) - \langle \bar{d}d\rangle m_o^2 + 4 \langle \bar{u}u\rangle m_o^2\Big],
\end{eqnarray}
and 
\begin{eqnarray}
\Gamma_{\mu\nu,1}&=&-\frac{4\langle\bar{u}u\rangle}{9 M^2}\Big[12 M^2 (-4 \langle \bar{d}d\rangle + \langle\bar{u}u\rangle) + 14 \langle \bar{d}d\rangle m_o^2 + 7 (2 \langle \bar{d}d\rangle - \langle\bar{u}u\rangle) m_o^2\Big],\nonumber\\
\Gamma_{\mu\nu,2}&=&-\frac{\langle g^2G^2\rangle}{72 M^2 \pi^2}\Big[8 M^2 (\langle \bar{d}d\rangle - 4 \langle \bar{u}u\rangle) - \langle \bar{d}d\rangle m_o^2 + 4 \langle \bar{u}u\rangle m_o^2\Big].
\end{eqnarray}
We should mention here that in these results we take the masses of light quarks as $m_u=0$ and $m_d=0$. After obtaining both sides, from these results we deduce two coupled sum rule equations. These equations are
\begin{eqnarray}
\lambda^*{}^2e^{-\frac{m^2}{M^2}}+\lambda_1^*{}^2e^{-\frac{m_1^2}{M^2}}+\lambda_2^*{}^2e^{-\frac{m_2^2}{M^2}}=\mathcal{\widehat B}\Pi_{\mu \nu,1}^{\mathrm{QCD}},\label{Eq1}
\end{eqnarray}
and
\begin{eqnarray}
m \lambda^*{}^2e^{-\frac{m^2}{M^2}}+m_1\lambda_1^*{}^2e^{-\frac{m_1^2}{M^2}}-m_2\lambda_2^*{}^2e^{-\frac{m_2^2}{M^2}}=\mathcal{\widehat B}\Pi_{\mu \nu,2}^{\mathrm{QCD}}.\label{Eq2}
\end{eqnarray}
As it can be seen from these results, there are six unknowns that we need to decide. These unknowns are $\lambda^*$, $\lambda_1^*$, $\lambda_2^*$, $m$, $m_1$ and $m_2$. Therefore these two equations are not enough to obtain these six unknowns and we need four more equations. Therefore we will consider first and second derivatives of these two equations, Eqs.~(\ref{Eq1}) and (\ref{Eq2}),  with respect to $(-\frac{1}{M^2})$ to obtain four more equations. After that, we solve all these six equations together to obtain the desired unknowns.

\section{Numerical Analyses}

The numerical analyses of the obtained sum rule equations require some input parameters, such as $m_o^2$, $\langle \bar{q}q\rangle$ and $\langle g^2G^2\rangle$ whose values are $m_{o}^2 =                            (0.8\pm0.1)$ $\mathrm{GeV}^2$ \cite{Belyaev:1982sa}, $\langle \bar{q}q \rangle (1\mbox{GeV})=(-0.24\pm 0.01)^3$ $\mathrm{GeV}^3$ \cite{Belyaev:1982sa} and $\langle g_s^2 G^2 \rangle=4\pi^2 (0.012\pm0.004)$ $~\mathrm{GeV}^4 $\cite{Belyaev:1982cd}. Along with these input parameters, there are two more auxiliary parameters that we need to fix to obtain reliable sum rules analyses. These parameters are the so called Borel parameter, $M^2$, and the threshold parameter, $s_0$. Their working regions are chosen so that the results have a moderate dependence on them. The working region of the Borel parameter is determined using the related criteria of the QCD sum rules: The lower limit for this parameter is determined considering the convergence of the OPE calculations.  To this end, we consider the ratio of higher dimensional terms in the OPE (contribution of the terms having dimensions seven, eight and nine) to the total contribution of the OPE and require that this ratio is $\leq 6~\%$.  To extract the upper limit of $M^2$ we consider the pole dominance and use the following criterion
\begin{eqnarray}
PC=\frac{\mathcal{\widehat B}\Pi_{\mu \nu,i}^{\mathrm{QCD}}(M^2,s_0)}{\mathcal{\widehat B}\Pi_{\mu \nu,i}^{\mathrm{QCD}}(M^2,\infty)}\geq\frac{1}{2},
\end{eqnarray}  
where $ PC $ denotes the pole contribution. As we include the three resonances simultaneously by choosing the threshold parameter around the third resonance, $ PC $ stands for contributions of three selected resonances.
These analyses result in the following interval for Borel parameter
\begin{eqnarray}
1.4~\mbox{GeV}^2\leq M^2\leq 1.6~\mbox{GeV}^2.
\end{eqnarray}
The threshold parameter is related to the energy of the excited states, therefore in our analyses we take its interval as
\begin{eqnarray}
1.75^2~\mbox{GeV}^2\leq s_0\leq 1.85^2~\mbox{GeV}^2.
\end{eqnarray}
Our analyses show that, with the above intervals,  we find $PC=57\%  $ in average. We also find that  the last three nonperturbative operators have a contribution of $6.4 \%   $ to the  total mass sum rules in average. Hence, the standard criteria of the method are nicely satisfied. 

\begin{figure}%
\centering
\subfigure[][]{%
\label{fig:ex3-a}%
\includegraphics[height=2in]{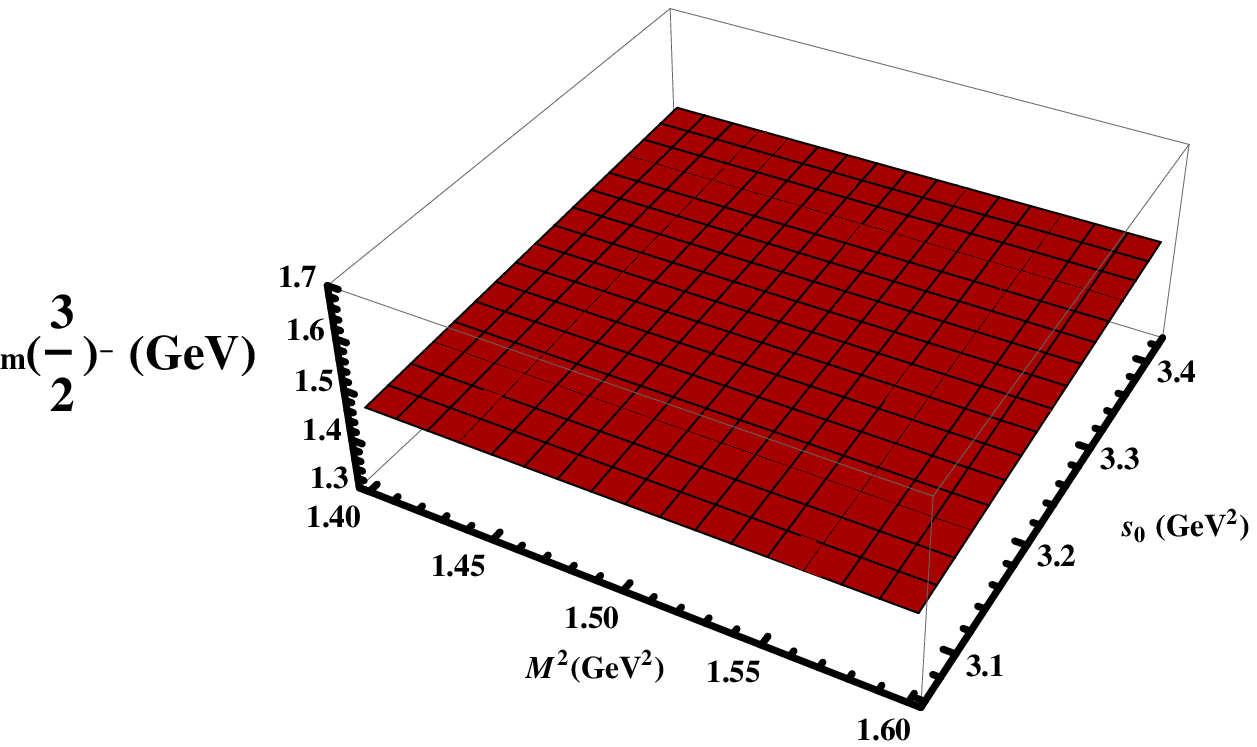}}%
\hspace{8pt}%
\subfigure[][]{%
\label{fig:ex3-b}%
\includegraphics[height=2in]{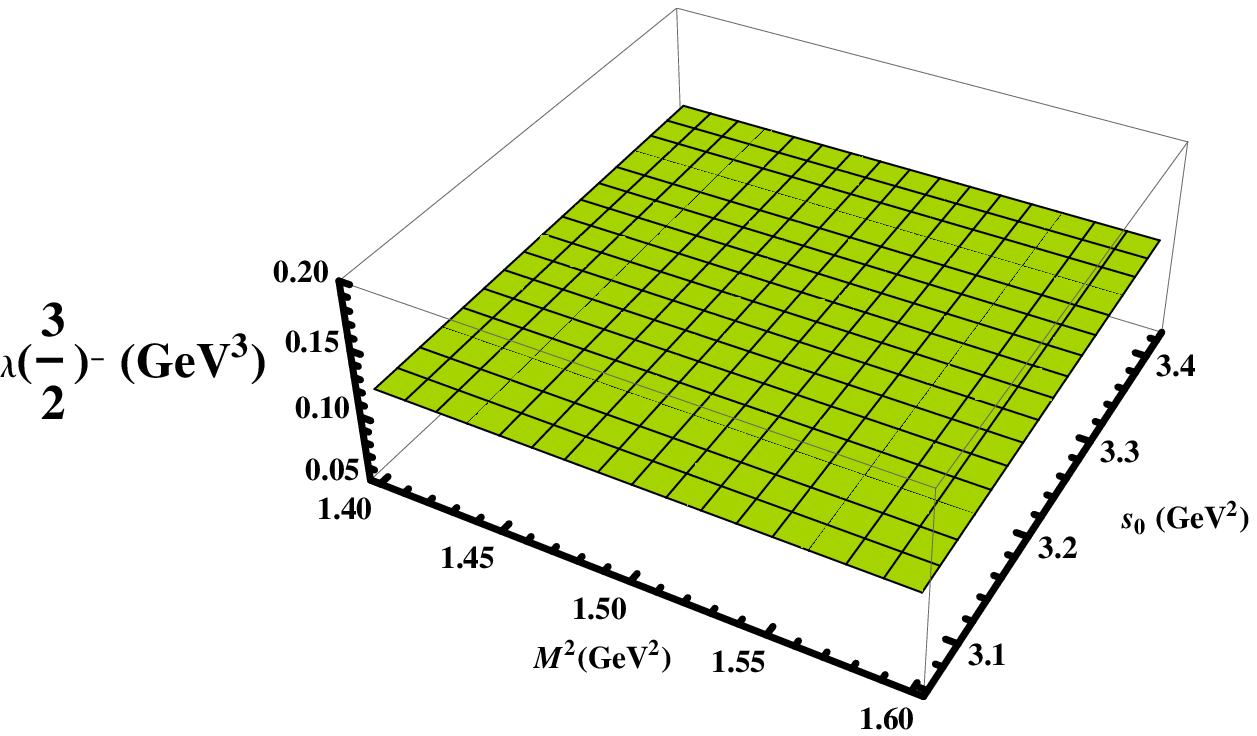}}%
\caption[A set of four subfigures.]{The mass and pole residue for the $N^*$ state obtained from QCD sum rule calculations as a function of $M^2$ and $s_0$ .:
\subref{fig:ex3-a} For mass; and,
\subref{fig:ex3-b} For pole residue.}%
\label{fig:massandpoleresidue}%
\end{figure}

To depict the behaviors of our results as a function of the auxiliary parameters, $M^2$ and $s_0$, we plot the mass and the pole residue graphs given in Fig.~\ref{fig:massandpoleresidue} obtained from our results for the $N^*$ state. The results show good stability against the variations of the threshold parameters, as desired.
With all these input parameters, we finally obtain the results for the masses and  pole residues of the considered states as in Table~\ref{results}. 
\begin{table}[]
\begin{tabular}{|c|c|c|}
\hline
      & Mass  (MeV)          & Pole Residue  (GeV$^3$)     \\ \hline\hline
$N^*(J^P=\frac{3}{2}^-)$ & $1505\pm 25 $ & $0.127\pm 0.004$  \\ \hline
$N_1^*(J^P=\frac{3}{2}^-)$ & $1701\pm62$   & $0.129\pm0.011$  \\ \hline
$N_2^*(J^P=\frac{3}{2}^+)$ & $1709\pm42$   & $0.111\pm 0.041$ \\ \hline
\end{tabular}
\caption{The masses and pole residues obtained from QCD sum rules.}
\label{results}
\end{table}
The errors in the results are due to the uncertainties of the input parameters and that of the auxiliary parameters. 

  At the end of this section we would like to mention that there are two ways to consider the resonances that all couple to the same interpolating current. The first one is to choose the threshold parameter around the energy of the first resonance and include into analyses only the first state. Then, by increasing the threshold the second resonance is included. For calculation of the parameters of the second resonance, the values of the parameters of the first resonance are considered as input parameters. Finally, using the parameters of the first two resonances and by increasing the threshold, the mass and residue of the third resonance are calculated. The second way, however, is to choose the threshold around the third resonance and include all the resonances into the analyses. In this method, the same number of sum rules with the number of unknowns are constructed to evaluate the parameters of the resonances under study. In the present study, we chose the second way as the resonances, especially the $N(1700)(3/2^-)$, and $N(1720)(3/2^+)$ resonances are very close to each other and it is very difficult to separate these states by increasing the threshold parameter step by step. In the second way, one may consider to include more resonances. For this, however, we will need more equations which can be constructed by applying higher order derivatives to find the entering unknowns. This leads to large uncertainties in the numerical results. Hence, we include into analyses only the first three resonances with spin $\frac{3}{2}$ and isospin $\frac{1}{2}$. 
  
\section{Summary and Conclusions}

In the present work we made analyses on the masses and pole residues for the low lying $N^*$ state with $J^P=\frac{3}{2}^{-}$ and its orbital and radial excitations.  In our analyses we used a powerful nonperturbative approach, namely QCD sum rules. We considered the resonances as ground state and its first orbital and radial excitations. Our mass results are in consistency, within the errors, with the experimental mass values of $N(1520)$ with mass $m_{N(1520)}=1505-1515$~MeV and $I(J^P)=\frac{1}{2}(\frac{3}{2}^-)$, $N(1700)$ with mass $m_{N(1700)}=1650-1750$~MeV and $I(J^P)=\frac{1}{2}(\frac{3}{2}^-)$ and $N(1720)$ with mass $m_{N(1720)}=1660-1690$~MeV and $I(J^P)=\frac{1}{2}(\frac{3}{2}^+)$~\cite{Tanabashi18}. 

We can also compare our findings with some other recent theoretical results. In Ref.~\cite{Giannini:2005ks} using G\"{u}rsey Radicati mass formula with two different sets of parameters the mass of the N(1520)D13 was obtained as $1543.7$~MeV and $1492.9$~MeV. Among these values the second one is in consistency with our result. In the same work the predictions for N(1700)D13 were given as $1658.6$~MeV and $1585.3$~MeV and for $N(1720)P13$ were given as $1651.4$~MeV and $1636.6$~MeV. These results are in accordance with ours within the errors.  The similar predictions were provided by Ref.~\cite{ZGhalenovi} via constituent quark model which were given as $m_{N(1520)}=1567.5$~MeV, $m_{N(1700)}=1657.5$~MeV and $m_{N(1720)}=1689.8$~MeV with errors presented as $3.09$\%, $2.50$\% and $2.04$\%, respectively. When we compare the results of Ref.~\cite{ZGhalenovi} with ours within their errors, it can be stated that all these results are consistent.  In Ref.~\cite{Santopinto:2014opa} the relativistic interacting quark-diquark model was used to obtain these masses and the results were presented as $m_{N(1520)}=1537$~MeV, $m_{N(1700)}=1625$~MeV and $m_{N(1720)}=1648$~MeV. In this work the prediction for $N(1700)$ is smaller than ours, but the other mass predictions are compatible. Using the semi-relativistic constituent three-quark model two values for the masses for each resonance were obtained with perturbative and approximative approaches, respectively, which were given in Ref.~\cite{MAslanzadeh} as $1511$~MeV and $1558$~MeV for $N(1520)$ resonance, $1667$~MeV and $1648$~MeV for $N(1700)$ resonance and $1735$~MeV and $1714$~MeV for $N(1720)$ resonance. The results obtained for $N(1520)$ and $N(1700)$ using perturbative approach are in agreement with our results, while for $N(1720)$ both the perturbative and approximative  approaches gave consistent results with that of our work.  The predictions for these resonances obtained using the hypercentral constituent quark model were given as $m_{N(1520)}=1535$~MeV and $m_{N(1720)}=1815$~MeV~\cite{Shah:2018ont} which are larger than our predictions. We collect all these information in the comparison in Table~\ref{results1}.

\begin{table}[]
\begin{tabular}{|c|c|c|c|c|c|c|c|}
\hline
      & Present Study         & Exp. \cite{Tanabashi18} &Ref.~\cite{Giannini:2005ks}&Ref.~\cite{ZGhalenovi}&Ref.~\cite{Santopinto:2014opa}&Ref.~\cite{MAslanzadeh} & Ref.~\cite{Shah:2018ont}\\ \hline\hline
$N^*(J^P=\frac{3}{2}^-)$ & $1505\pm 25 $ &$ 1505-1515 $ &$ 1543.7/1492.9 $&$1567.5  $& $ 1537 $&$1511/1558  $ &$ 1535 $\\ \hline
$N_1^*(J^P=\frac{3}{2}^-)$ & $1701\pm62$   &$ 1650-1750 $ &$1658.6/1585.3 $&$ 1657.5 $& $1625  $&$ 1667/1648 $&$ -$ \\ \hline
$N_2^*(J^P=\frac{3}{2}^+)$ & $1709\pm42$   &  $ 1660-1690 $&$1651.4/1636.6 $&$ 1689.8 $&$  1648$&$ 1735/1714 $&$1815  $\\ \hline
\end{tabular}
\caption{The masses of the considered resonances in MeV compared to the experimental data and other theoretical predictions.}
\label{results1}
\end{table}

Here we should put emphasis on one point. As we stated, considering the errors of our results, we obtained consistent results with the experimental observations and we could fix the central values of the masses of the considered excited nucleons. However, as it can be seen from our predictions for $N(1700)$  and $N(1720)$, the uncertainties in our results make their mass values overlap. This situation is also similar to their experimental observations, in which the mass values are given as some intervals. Because of this situation, it is necessary to support our results by means of other investigations on their  properties such as their strong, weak or electromagnetic decays. The usage of the obtained predictions of the present study in such analyses and comparison of their outcomes with both existing experimental and theoretical findings help us make their properties much clear and fix their masses unambiguously. 

To sum up, the results obtained in this work are important in understanding the nature of these nucleon resonances and may shed light on their underlying effective degrees of freedom. And also they are useful tools in the analyses of their decay mechanisms. Besides, a deeper understanding of the nature of these resonances improves our knowledge of strong interaction at low energy and may give insight into the missing resonance problem.


\label{sec:Num}


\end{document}